# Can Hong-Ou-Mandel type quantum correlation be a test tool for quantum entanglement?

Byoung S. Ham

School of Electrical Engineering and Computer Science, Gwangju Institute of Science and Technology
123 Chumdangwagi-ro, Buk-gu, Gwangju 61005, South Korea
(Submitted on June 25, 2022; bham@gist.ac.kr)

**Abstract**
Quantum technologies based on the particle nature of a photon has been progressed over the last several decades, where the fundamental quantum feature of entanglement has been tested by Hong-Ou-Mandel (HOM)-type anticorrelation as well as Bell-type nonlocal correlation. Mutually exclusive quantum natures of the wave-particle duality of a single photon have been intensively investigated to understand the fundamental physics of 'mysterious' quantum nature. Here, we revisit the HOM-type quantum correlation to answer the question whether the HOM-type anticorrelation can be a test tool for quantum entanglement. For this, a pair of spontaneous parametric down converted photons is tested for the anticorrelation of HOM effects, where the SPDC-generated photon pair is not in an entangled state.

**Introduction**
A corpuscular nature of light has been accepted since the photoelectric effect was observed by Einstein in 1905. Entanglement [1] between paired photons has been understood in terms of the particle nature of a photon, where a photon, on the other hand, also has a wave nature according to the Copenhagen interpretation [2-4]. Based on the fundamental understanding of the wave-particle duality in quantum physics [2], these two mutually exclusive natures of a photon are complementary [5,6]. Thus, one nature (energy) of a photon cannot be simultaneously appeared alongside the other (phase). In other words, specifying a photon's energy in a Fock state results in vagueness for its phase information. This fundamental physics of the energy-phase uncertainty relationship, however, does not apply to paired photons. Without understanding of the phase basis in such a coupled system, quantum features become mysterious and probabilistic. In that sense, a clear definition of the classicality and quantumness of a paired system is prerequisite to understand quantum correlation. The understanding of quantum correlations between two bipartite photons has recently been revisited to account for the wave nature with a mutual coherence basis, even though individual photons have no specific phase information [7,8]. Here, we study a paired photon system acting on a beam splitter (BS) to understand the fundamental quantum characteristics of Hong-Ou-Mandel (HOM) effects [9], whether the HOM effect can be a test tool for quantum entanglement.

Entanglement between paired photons or atoms is known as a weird quantum phenomenon, where specifying the mutual phase relationship has never been considered with respect to the probabilistic nature of quantum mechanics [10-20]. Thus, post-measurement techniques have been developed for the probabilistic nature of quantum correlation. Recently, a completely different approach has been suggested to understand the fundamental physics of anticorrelation in HOM effects [7] and Franson-type nonlocal correlation [8,21]. In regard to quantum resources of light, $\chi^{(2)}$ nonlinear optics-based spontaneous parametric down conversion (SPDC) process has been used for entangled photon-pair generations [22]. Due to the vagueness of phase information between SPDC-generated entangled photons, however, the post-measurement-based quantum techniques show probabilistic quantum characteristics. Here, the anticorrelation of HOM effects is investigated whether the HOM effect can be a test tool for quantum entanglement. For this, pure coherence approach is used to comply with the interferometric BS system [23], resulting in a complete and definite causality. Unlike the common understanding of the probabilistic nature of a photon pair on a BS [9-20], mutual coherence between paired photons results in a deterministic nature [7]. As a result, the solution of the anticorrelation of HOM effects on a BS is coherently driven from the BS system applied by the quantum particles of SPDC-generated photon pair. This photon pair is, however, provided in a non-entangled state to answer the raised question.

**Analysis**

Figure 1 shows a basic quantum measurement scheme of paired photons impinging on a BS [10]. The paired signal and idler photons are not entangled but frequency/momentum correlated directly from the SPDC process [22]. For the SPDC-generated photon pair, a standard particle nature-based analysis cannot show a deterministic quantum feature due to no definite phase relationship between the paired photons. As already observed by many research groups over the last several decades, the observed anticorrelation of the HOM effects [9] has been used as a proof of quantum entanglement between the paired photons [24]. In Fig. 1, we analyze whether such an understanding of the HOM effects is a sufficient and necessary condition for the test of quantum entanglement such as in Bell inequality violation. For this, we simply add individual and random phase information to both input photons to use their mutual (difference) phase information for the same coincidence measurements. This mutual phase relationship between the paired photons does not violate quantum mechanics [1,2,22]. Thus, the present analysis becomes classical based on the wave nature of quantum mechanics. As a result, a definite solution of the HOM effect is sought for a quantum pair whose entanglement is intentionally avoided to answer the question.

One of the most fundamental quantum features is basis randomness via quantum superposition between them [25]. In a typical Young's double slit or a Mach Zehnder interferometer (MZI), the fundamental phase-basis set is denoted by $\theta \in \{0, \pi\}$, where each basis relates to a pure state of a particle in an interferometric system. The origin of this phase bases of a single photon is in the harmonic oscillation defined by Maxwell's wave equation. Thus, the wave nature of quantum mechanics results in a single-photon self-interference via basis randomness [26]. In Fig. 1(a), the fundamental phase-basis set of each input photon impinging on a BS is denoted by $\varphi \in \{0, \pi\}$. The definition of anticorrelation of the HOM effects is zero coincidence measurement for two output photons [9]. Here, the measurement randomness in HOM effects is limited to each output port under the coincidence detection scheme, resulting in $\langle I_A \rangle = \langle I_B \rangle = I_0$, where $I_0$ is the single photon intensity. The amplitude probability of $E_j$ of each photon is the fundamental resource of the randomness known as Born's rule [27-29]. According to the Born rule test [28,29], Fig. 1 is a two-input-two-output system, where the results cannot be distinguished from those of continuous waves.

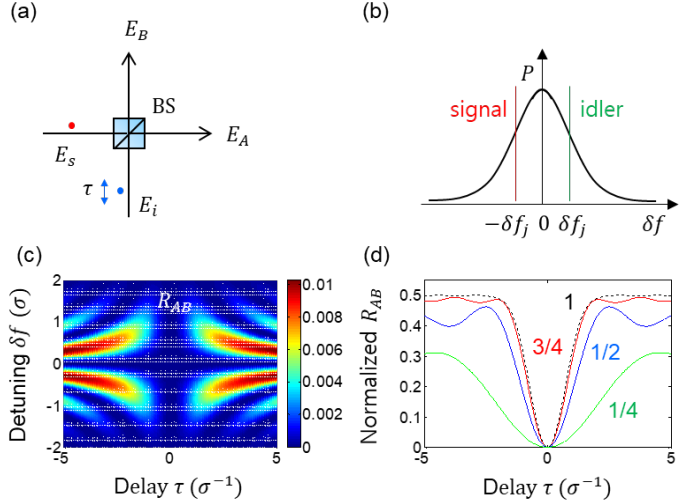

Fig. 1. SPDC generated photon pair interactions on a beam splitter for anticorrelation. (a) Schematic of anticorrealtion of HOM effects. (b) Spectral distribution of SPDC generated photon pairs. (c) and (d) Numerical calculations for equation (4). In (c), the detuning $\delta f$ values are averaged and normalized with respect to the delay τ. The numerical values in (d) indicate the ratio of reduced bandwidth to the dotted curve of (c): dotted (1), red (3/4), blue (1/2), green (1/4). BS: a 50/50 nonpolarizing beam splitter. The SPDC photon-pair distribution is Gaussian with standard deviation σ.

For paired two input photons impinging on a BS in Fig. 1(a), each photon's amplitude can be described as $E_j(\mathrm{r,t}) = E_0 e^{i\varphi_j(r,t)}$, where $E_0$ ($\varphi_j$) is amplitude (phase) of a single photon. Although a definite phase of each single photon in Fig. 1(a) violates quantum mechanics, a relative phase $\delta\varphi$ between them ($E_s$ and $E_i$) does not.

The time delay $\tau$ between $E_s$ and $E_i$ is with respect to the coincidence on the BS. Using the BS matrix representation [23], the following output intensities are obtained:

$$I_A = I_0(1 - sin\delta\varphi), \quad (1)$$
$$I_B = I_0(1 + sin\delta\varphi), \quad (2)$$

where $E_A = (E_s + iE_i)$, $E_B = (E_s - iE_i)$, $I_0 = E_0 E_0^*$, and $\delta\varphi = \varphi_s(\delta,\tau) - \varphi_i(\delta,\tau)$. Here, the initially given global phase in each pair does not affect Eqs. (1) and (2). For the uniform output intensity, however, a pair of opposite phase difference $\pm\delta\varphi$ is needed. Such a relation is satisfied by the intrinsic property of the SPDC-generated photon pairs, as shown in Fig. 1(b) [22], resulting in uniform average values, $\langle I_A \rangle = \langle I_B \rangle = I_0$. For this, the SPDC output photons does not satisfy entanglement defined by $|\Psi\rangle_{ent} = \frac{1}{\sqrt{2}}(|signal\rangle_1|idler\rangle_2 + e^{i\psi}|signal\rangle_2|idler\rangle_1)$, unless quantum superposition between the signal and idler photons is additionally conducted, where the subscript relates to each path of an input photon in Fig. 1(a) [22]. Thus, Fig. 1(a) is for either the first or second correlated term only in $|\Psi\rangle_{ent}$. More importantly, the random global phase given to each photon pair in the SPDC process has nothing to do with measurements if there is a fixed phase relationship between the paired photons [7]. In that sense, Eqs. (1) and (2) basically satisfy the Young's double double-slit experiments governed by coherence optics. For the condition of $\pm\delta\varphi$, the symmetric frequency distribution in SPDC-generated photon pairs in Fig. 1(c) is the origin of the local randomness observed in each output port of Fig. 1(a). Here, the quantum operator approach in conventional understanding based on the particle nature of a photon assumes the same phase relation between two input photons applied for the BS matrix, where the $\frac{\pi}{2}$ phase difference between paired photons cannot be driven [30].

The normalized coincidence measurement between $I_A$ and $I_B$ for individual photon pairs, however, must result in definite phase relationship, resulting in the anticorrealtion of the HOM effect [9]. Using the phase correlation between SPDC-generated paired photons with a definite and fixed phase difference, the following relation can be obtained from Eqs. (1) and (2):

$$\langle R_{AB}(\tau) \rangle = I_0^2 cos^2(\delta\varphi), \quad (3)$$

where Eq. (3) never satisfies anticorrelation of the HOM effect. To work with anti-correlation, however, modified Eqs. (1) and (2) with a $\pm\frac{\pi}{2}$ phase shift in $\delta\varphi$ is needed, resulting in: $\langle R_{AB}(\tau) \rangle = I_0^2(1 \mp sin\delta\varphi')(1 \pm sin\delta\varphi')$, where $\delta\varphi' = \delta\varphi \pm \frac{\pi}{2}$. This relationship is the uniqueness of the SPDC-generated photon pairs that cannot be achieved in coherent photons from a laser, resulting in:

$$\langle R_{AB}(\tau) \rangle = I_0^2 sin^2(\delta\varphi). \quad (4)$$

Due to the squared sine function, coincidence measurements show the same result for the symmetric detuning $\pm\delta\varphi$. Thus, the opposite phase relation with a $\pm\frac{\pi}{2}$ phase shift between the paired photons in SPDC is key to understanding the quantum feature of HOM effects. The ensemble average of the coincidence measurement is $\langle R_{AB}(\tau) \rangle = \frac{1}{N}\sum_j^N sin^2\delta f_j \tau$. As $\tau$ increases between two input photons, $\delta\varphi$ ($= \delta f_j \tau$) increases, too, as shown in Figs. 1(c) and (d). As a result, $R_{AB}(0) < \langle R_{AB}(\tau) \rangle < \frac{1}{2}$ is obtained, where the upper limit of $\langle R_{AB}(\tau) \rangle = \frac{1}{2}$ in Fig. 1(d) becomes the lower bound of the classical feature, representing individual particles with no phase relationship as defined by Bell [9].

Numerical calculations of the normalized coincidence measurements for Eq. (4) are shown in Figs. 1(c) and (d). As $\tau$ increases, the perfect anticorrelation of the HOM effects at $\tau = 0$ degrades as shown in Fig. 1(d) (see the dotted curve for full bandwidth). This $\tau$-dependent correlation loss saturates at $\tau \geq 2\sigma^{-1}$, where $\sigma$ is the standard deviation of the photon spectral distribution. As expected in Fig. 1(c), spectrally filtered input photons alleviate the correlation loss, as shown by the colored curves in Fig. 1(d) due to enhanced coherence by $\delta f \tau$. The lack of λ-dependent fringe observed in the HOM effects is because of the $\tau$-resulting coherence washout in an ensemble. Thus, the anticorrelation of the HOM effects on a BS is no longer mysterious but deterministic according to the phase relation between paired photons. As an extreme case of monochromatic light pair satisfying $\pm\delta\varphi$ and $\frac{\pi}{2}$ phase difference, the coincidence detection has no meaning, and the wave nature dominates in the same quantum feature of the anticorrelation.

**Conclusion**

Coherence characteristics on a BS were investigated for SPDC-generated photon pairs with no entanglement. In the wave nature-based analysis of the HOM effects [9], a deterministic solution was driven for the SPDC-generated paired photons. Furthermore, numerical simulations were tested for coincidence measurements between two output photons from the BS, where bandwidth-dependent coherence degree directly affects the degree of quantum correlation in the HOM effects. When the photon bandwidth was spectrally reduced, such τ-dependent coherence washout became alleviated.; As a result, a correlation modulation fringe was appeared, as observed in Ref. [31]. Unlike conventional understanding, thus, the anticorrelation of HOM effects was understood as a deterministic coherence feature based on mutual phase relationship of the ensemble of SPDC-generated photon pairs. Thus, the anticorrelation of the HOM effects cannot be used for a test tool of quantum entanglement. On the other words, anticorrelation of the HOM effects is not sufficient to prove quantum features. The Bell measurement of inseparable intensity products via coincidence detection is different from the HOM effects, where the inseparability between locally measured intensity products proves quantum entanglement.

**Acknowledgments**

This work was supported by the ICT R&D program of MSIT/IITP (2021-0-01810), development of elemental technologies for ultrasecure quantum internet and the GIST Research Project in 2022.